\documentclass[AMA,STIX1COL]{WileyNJD-v2}

\articletype{Article Type}%

\received{}
\revised{}
\accepted{}
        
\begin{document}

\title{From Distributed Machine Learning To Federated Learning: In The View Of Data Privacy And Security}

\author[1]{Sheng Shen}
\author[1]{Tianqing Zhu*}
\author[2]{Di Wu}
\author[1]{Wei Wang}
\author[1]{Wanlei Zhou}

\authormark{Sheng Shen \textsc{et al}}

\address[1]{\orgdiv{University of Technology Sydney}, \orgname{Center for Cyber Security and Privacy, School of Computer Science}, \orgaddress{\state{NSW}, \country{Australia}}}
\address[2]{\orgdiv{University of Technology Sydney}, \orgname{School of Computer Science}, \orgaddress{\state{NSW}, \country{Australia}}}

\corres{*Tianqing Zhu, \email{Tianqing.zhu@uts.edu.au}}

\presentaddress{15 Broadway, Ultimo, NSW 2007, Australia}

\abstract[Summary]{Federated learning is an improved version of distributed machine learning that further offloads operations which would usually be performed by a central server. The server becomes more like an assistant coordinating clients to work together rather than micro-managing the workforce as in traditional DML. One of the greatest advantages of federated learning is the additional privacy and security guarantees it affords. Federated learning architecture relies on smart devices, such as smartphones and IoT sensors, that collect and process their own data, so sensitive information never has to leave the client device. Rather, clients train a sub-model locally and send an encrypted update to the central server for aggregation into the global model. These strong privacy guarantees make federated learning an attractive choice in a world where data breaches and information theft are common and serious threats. This survey outlines the landscape and latest developments in data privacy and security for federated learning. We identify the different mechanisms used to provide privacy and security, such as differential privacy, secure multi-party computation and secure aggregation. We also survey the current attack models, identifying the areas of vulnerability and the strategies adversaries use to penetrate federated systems. The survey concludes with a discussion on the open challenges and potential directions of future work in this increasingly popular learning paradigm.}

\keywords{federated learning, data privacy, security, distributed machine learning}

\maketitle

\section{Introduction}
Since Google \cite{konevcny2016federated, konevcny2017federated, mcmahan2016communication} first proposed the concept, federated learning has become an intriguing topic in privacy-preserving machine learning. Sometimes called collaborative learning, federated learning mobilizes multiple clients, such as computers, processing devices or smart sensors, coordinated by a central server to collaboratively train a machine learning model. Google's idea was to distribute training sets across across multiple devices and have each contribute to building the model, all while preventing data leaks \cite{yang2019federated}. Clients train part of the model and upload partial updates to a central server, which then averages those updates and applies the result to the global model. The scheme considers the size of the training data, the computing resources required and the data privacy and security concerns. Each client is an isolated “data island”, and data never leaves the island. Once used to train the model, the only ’ship’ to leave is a model update to the central server. Plus, no island has all the data. Due to the limitation of a data island on the size and the characteristic, federated learning framework ideally benefits more to clients to collaboratively train a machine learning model using their data in security. The result is a more effective model that is insensitive to the raw data of others and, thus, federated learning has proven to be particularly attractive to governments, hospitals and financial institutions.

\begin{figure}[htbp]
\centering
\begin{minipage}[t]{0.48\textwidth}
\centering
\includegraphics[width=7.8cm]{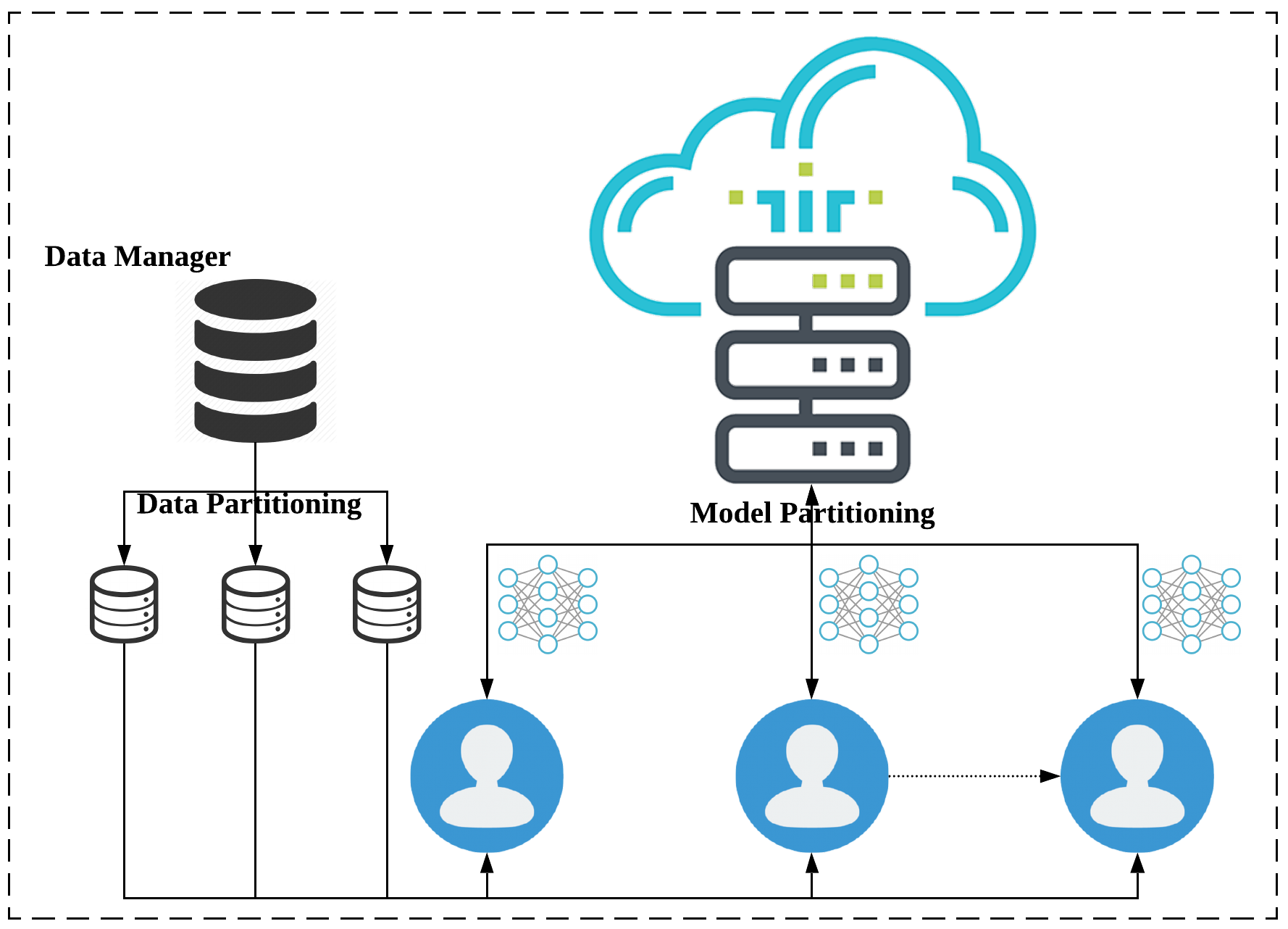}
\caption{A basic DML system}
\label{Fig-DMLOverview}
\end{minipage}
\begin{minipage}[t]{0.48\textwidth}
\centering
\includegraphics[width=7.8cm]{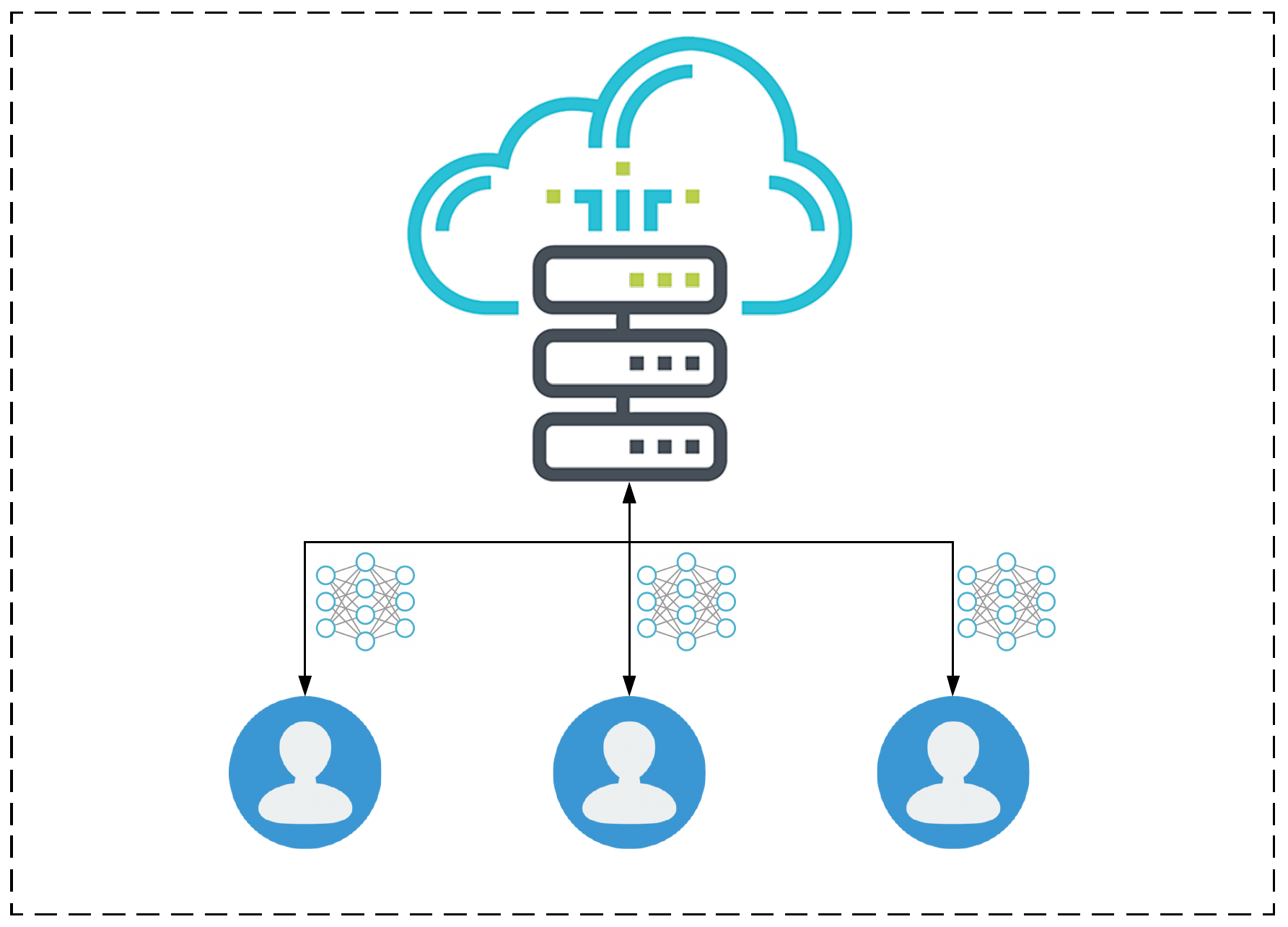}
\caption{A basic federated learning system}
\label{Fig-FLOverview}
\end{minipage}
\end{figure}

Federated learning is a specific category of distributed machine learning (DML) \cite{bonawitz2019towards} that further decentralizes operations that would usually be performed by the server. The server becomes more like an assistant that coordinates clients to work together instead of micro-managing schedules as in traditional DML. \textbf{FIGURE \ref{Fig-DMLOverview}} shows the classic DML framework, which includes a central server, some clients and a data manager. The central server can act as the data manager, or the data can be managed by a third party storage system under the server’s control. Together, the server and the data manager use optimization strategies to partition the training data into many subsets and the model into many parts and then disseminate learning tasks to the clients. Note that a key difference between DML and federated learning is that, in DML, one client may ask other clients to transfer their training data if needed to meet their own learning prerequisites or conditions.

In comparison, \textbf{FIGURE \ref{Fig-FLOverview}} illustrates a typical federated learning system. First, a central server publishes a machine learning task and selects clients to participate in each epoch of the training process. Then it sends the model and relevant sources to the clients and waits for their training results. Clients train the model with the data on their device and return an update of the model parameters or gradients to the server. The server then aggregates those details and updates the 'master' model for the next training epoch. There two key advantages with this type of learning scheme: reduced computational and communications overhead and better privacy. The details of DML and federated learning will be introduced in the next section.

In fact, federated learning can incorporate many privacy preserving and security mechanisms across the entire system -- from the collaborative training process to aggregating updates at the server. For instance, differential privacy (DP) \cite{dwork2006calibrating} and local DP \cite{kairouz2016extremal} can guarantee that both the training data and the updates remain private at the numeric level. Secure aggregation protocols on the server side, consisting of secure multi-party computation, secret sharing and homomorphic encryption, can perturb the updates to guarantee model security during transfer and aggregation. 

However, although federated learning has made huge improvements to the privacy and security of machine learning models and all their associated processes, it is not a perfect solution. As with many new computing paradigms, federated learning is attracting its share of attention from adversaries with malicious intent. These adversaries might be internal agents participating in the training process who can influence the model updates, or they may be external agents that can only observe the learning and update process but are still able to make inferences that comprise a person's privacy. Therefore, federated learning is still vulnerable to information leaks and many other types of attacks, such as backdoor attacks, model poisoning and inference attacks. A detailed survey of these adversarial models appears in a later section. Although some comprehensive surveys on federated learning have been published in the past three years, most focus on reviewing the systems and applications of federated learning. Few mention privacy preserving and security \cite{baruch2019little, yang2019federated}, and none go into detail. Hence, the focus of this survey is on the privacy preserving and security aspects of federated learning, including privacy concerns, techniques for protecting privacy and securing assets, adversarial models, the challenges the field faces today, and future directions for advancement.

In the next section, we compare DML and federated learning from the perspective of privacy preservation. Then, in Section 3, we provide an in-depth analysis of the current mechanisms used in federated learning to provide privacy and security guarantees. Section 4 presents some of the most common and effective attack models against federated learning. We demonstrate the ways in which federated learning is still vulnerable to some methods of attack along with some possible defense strategies. Promising fields and applications for federated learning are outlined in Section 5, followed by the conclusion and future directions of research in Section 6.

\section{From Distributed Machine Learning To Federated Learning}
\subsection{Distributed Machine Learning}
DML is a combination of distributed computing and machine learning,. DML has a very fast learning speed, so it is widely used for tasks with large-scale data or model parameters. The principle is to partition the training data and the model and have a parameter server devise and co-ordinate a schedule of multiple clients which learn each partition as a sub-task. All clients learn their allocated sub-task in parallel and, when all clients have completed their work, the parameter server aggregates the sub-models together and generates a complete model through scheduled aggregation algorithms. To train the model more effectively, the sub-models should simultaneously match the sub-tasks in order to train the model more effectively. Obviously, this process relies heavily on good communication between the server and the clients. However, it is important to strike a balance between the learning and communication costs because, with large scale data, resource constraints on storage, bandwidth and time can present real problems. As such, with DML frameworks, proper scheduling is vital to efficient performance.

\subsubsection{The Structure and Data Flows of Distributed Machine Learning}

\begin{figure*}[htpb]
\centering
\includegraphics[width = \textwidth]{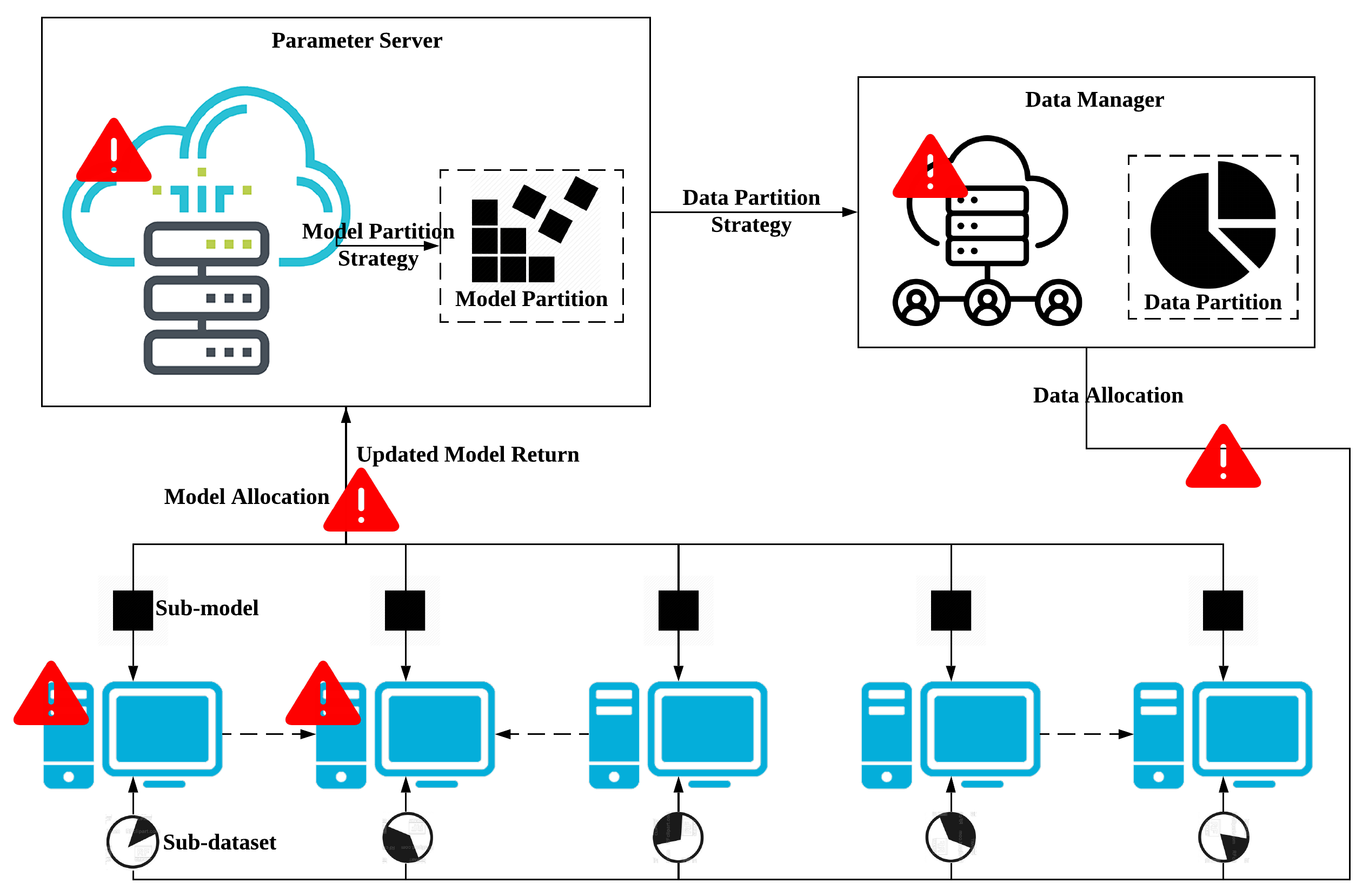}
\caption{The architecture and data flow of distributed machine learning}
\label{Fig-DML}
\end{figure*}

\textbf{FIGURE \ref{Fig-DML}} shows the basic architecture of DML. The parameter server is central to the system. It is responsible for: scheduling; partitioning the model and the training data; allocating sub-tasks to clients; and aggregating the sub-models. The data manager could be the server or a third-party storage device obeying the server’s partition strategy. Clients can complete sub-tasks independently but, if a sub-tasks has prerequisite or follow-on tasks, they can also communicate with other clients to get the data they need. 

From the perspective of data flow, the parameter server must obviously have to access to the whole dataset to be able to create the partitions no matter whether the data is managed by the central server or a third-party. Notably, this data flow is one way; the partitioned data does not get sent back to the server; only the sub-model update does.

\subsubsection{Privacy and Security Problems and Adversaries in DML}
The biggest, but not the only, vulnerability of DML is the amount of communication needed between the parameter server and the clients. Like the highway robberies of old, privacy protection is at its weakest when data is in transit. Therefore, the more communication, the more opportunities there are for attack. The danger alerts in \textbf{FIGURE \ref{Fig-DML}} indicate likely points of intrusion by adversaries. They include, the parameter server as the brain of the system, the data manager, the individual clients who may or may not be secure, as well as any time data is transmitted from one device to another. If an attack is successful, the amount of data leaked depends on the location of the attack. Violating a client may only net one or two data partitions but successfully penetrating the parameter or data server may yield the entire database or the entire model.

\textbf{TABLE \ref{table1}} summarizes the types of adversaries and their targets against DML schemes. Spectators can only observe the algorithms, models and training process. These adversaries are most likely curious about the training data and model but cannot affect the learning process. Conversely , participant adversaries can do quite a lot more damage. For instance, a malicious parameter server could wreak havoc because of its strong and centralized power, whereas the damage done by an adversarial clients is more contained. Hence, higher-level devices in the architecture are more attractive to adversaries.

\begin{table*}[ht]
    \centering
    \begin{tabularx}{\textwidth}{ll}
        \centering
        \begin{minipage}{\textwidth}
            \includegraphics[width = \linewidth]{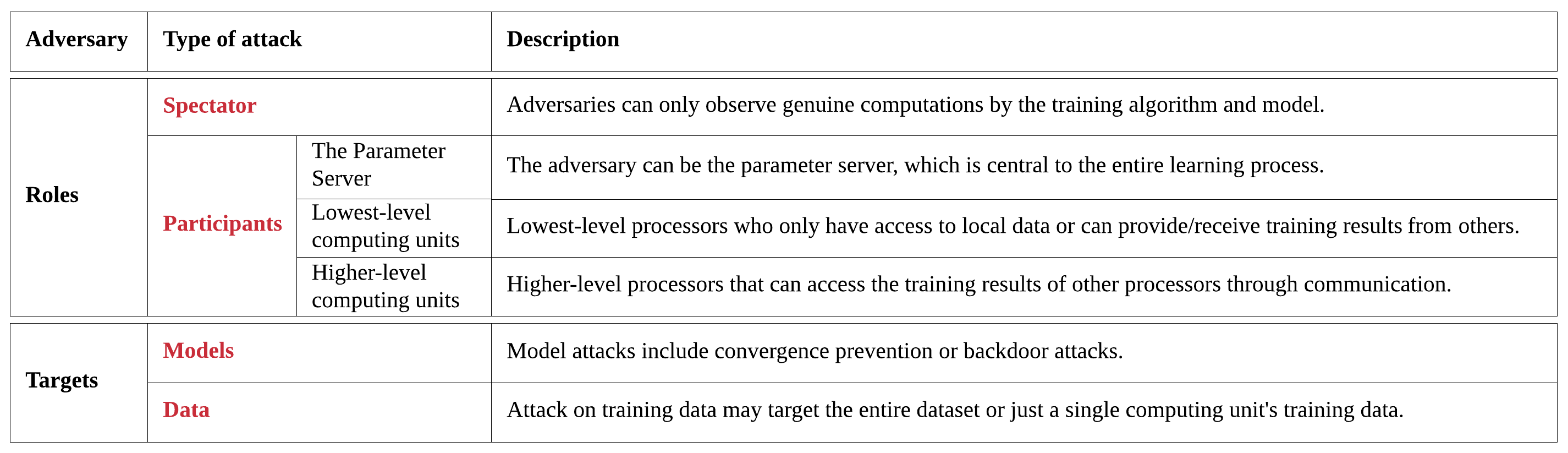}
        \end{minipage}
    \end{tabularx}
    \caption{Type of adversaries and their targets against DML systems}
    \label{table1}
\end{table*}

\subsection{Federated Learning}
Federated Learning is a specific type of DML, designed to overcome some of the privacy and security issues with classic DML architecture.

\subsubsection{Basic Structures and Data Flow}

\begin{figure*}[htpb]
\centering
\includegraphics[width = \textwidth]{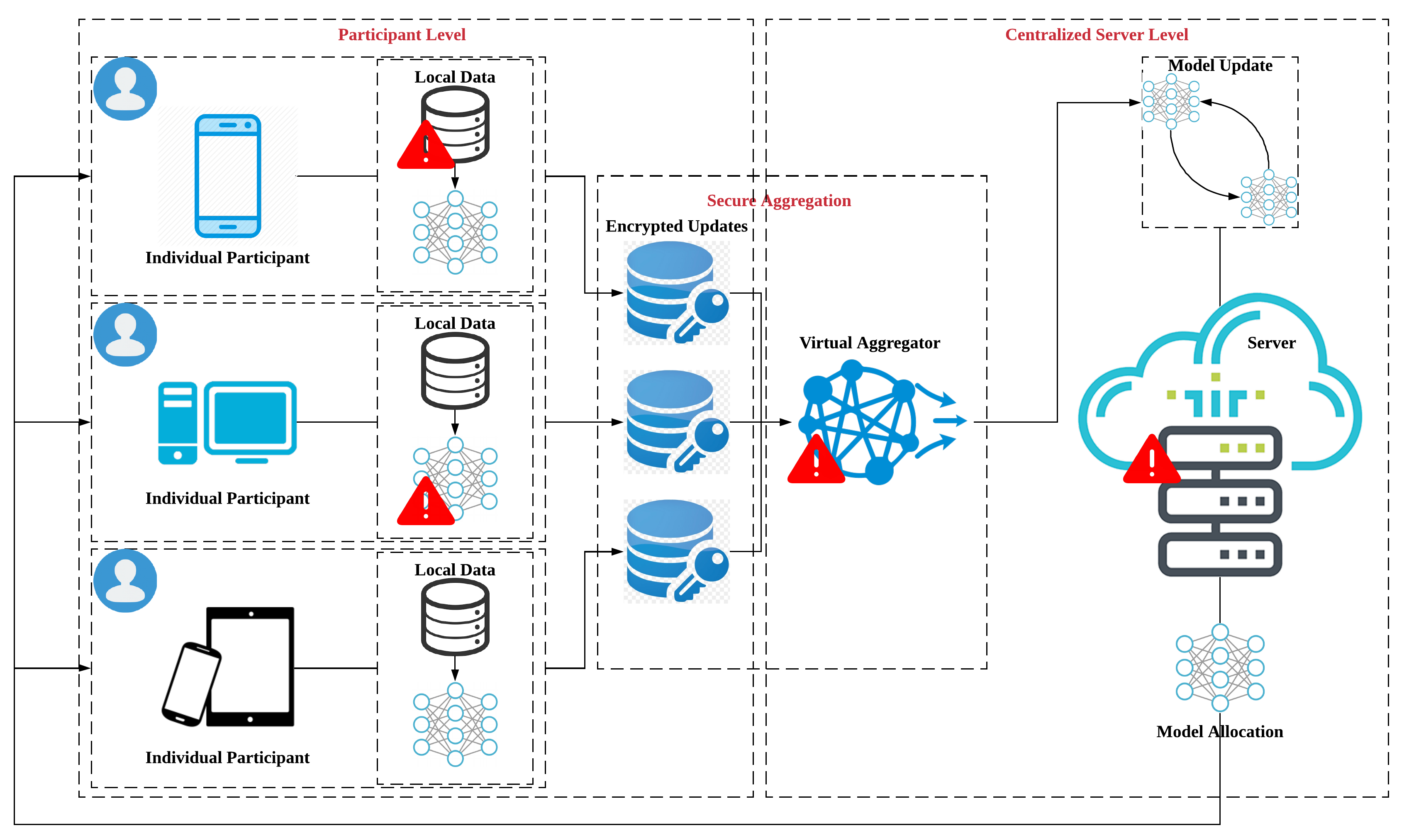}
\caption{Federate Learning Structure And Data Flow}
\label{Fig-FL}
\end{figure*}

The basic architecture of federated learning including its data flows is illustrated  in \textbf{FIGURE \ref{Fig-FL}}. As shown, there are some similarities and some differences between DML and federated learning. Like traditional DML, there is a central server, which is responsible for the overall control and management of training a global model and some clients who receive training sub-tasks from the central server. There are two key differences however: a)  Instead of each client working individually on their own piece of the model, in federated learning, all selected clients work on the same training task in each epoch; and b) The clients in a federated learning system are typically devices like smart phones, tablets, and sensors that are able to capture or collect information as opposed to desktop computers or routers. Therefore, because the training data is gathered, stored and used at the client level, the only information that ever needs to be transmitted is the model updates.

The learning procedure is relatively straightforward. In each training epoch, the server allocates a training task and computing resources to any client that is ready to learn, then it transmits the current model. The client trains the model with its own local data and sends the updated parameters as encrypted training results back to an aggregator for compilation. As such, there is greater data privacy because there is no need to transmit sensitive information, and encrypting the updated parameters before sending them to aggregators increases security over the models.

The aggregators, also controlled by the central server, average the parameter updates. There are two types of aggregators: master and temporary. Master aggregators manage the number of training epochs and generate an appropriate number of “temporary” aggregators for each epoch to consolidate the training results. These temporary aggregators do not store any permanent information, and all aggregators follow what is called a “secure aggregation protocol”, which means encrypted data can be processed and compiled without knowing the true data. The master aggregators then fully aggregate the results from the temporary aggregators and deliver the results to the central server that updates the model. The server then schedules the next training task and starts a new training epoch.

\subsubsection{Privacy And Security Problems And Adversaries In Federated Learning}

Even though federated learning was designed improve privacy and security, these frameworks still have vulnerabilities and security risks. Again, the danger signs in \textbf{FIGURE \ref{Fig-FL}} identify potential attack targets. First, the raw data on the client devices  is an attractive target for adversaries. Even though these data are never transmitted, they are still open to inference attacks without proper privacy guarantees at the device level. Second, the master model is a very valuable prize, which could be targeted in either a master aggregrator or the central server. The different types of adversaries and their potential attack targets are summarized in \textbf{TABLE \ref{table2}}.

\begin{table*}[ht]
    \centering
    \begin{tabularx}{\linewidth}{ll}
        \begin{minipage}{\textwidth}
             \includegraphics[width = \linewidth]{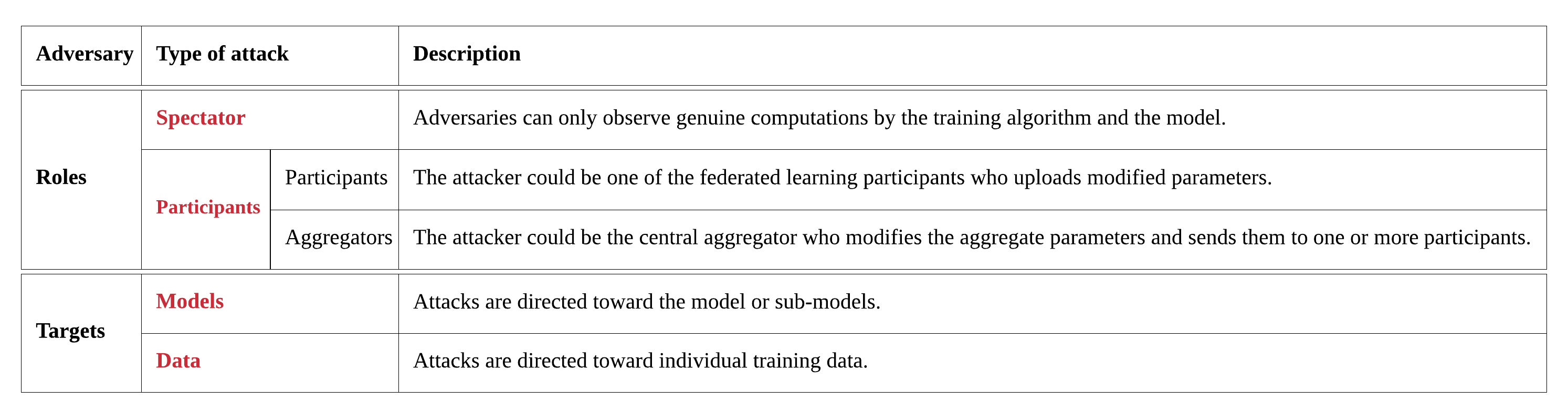}
        \end{minipage}
    \end{tabularx}
    \caption{Type of adversaries and their targets against federated learning systems}
    \label{table2}
\end{table*}

As well as adversaries targets, we also cover potential roles of adversaries in \textbf{TABLE \ref{table2}}. In federated learning, an adversary can be either a spectator or a participant. Malicious spectators of the model’s training process can observe, but they cannot affect model performance, so the vulnerability here is one of an inference attack. The target might be either the model’s parameters or an attempt to glean sensitive information from the data. Malicious participants, however, can both observe and change individual updates, while malicious aggregators can observe the global parameter updates and control the results of averaging.

\subsection{From DML to Federated Learning}

Federated learning offers some significant improvements over DML in terms of both security and efficiency. Although there is still work to be done to make federated learning near to a completely secure system, this scheme seems the inevitable future of decentralized learning. What follows is a summary of the major advancements from DML to federated learning to date.

\begin{enumerate}
    \item Federated learning uses smartphones, tablets and sensors as clients instead of computers and other non-data gathering devices. This means data is collected, used, and stored autonomously. Data does not need to be transferred around the system and no data manager is required. Less communication means less risk of data leaks and greater data privacy. In fact, data privacy protection is a fundamental property of federated learning. As a comparison between Figures \ref{Fig-DML} and \ref{Fig-FL} show, federated learning is not subject to direct privacy leaks during communication.
    
    \item Unlike DML, which allocates model training to clients in a piecemeal fashion, federated learning operates on the principle of collaboration where selected clients work on the same training task in parallel. Both systems still transmit parameter updates between the client and various servers. However, the collaborative nature of this approach means there is no need for a chain or hierarchy of clients and no reason for any client-to-client transfers. This further reduces the risk of data leaks.
    
    \item Third, federated learning involves less communication than DML which reduces the opportunity for attack risks. As mentioned, the central server in DML allocates both data and model partitions to clients by communication. There may also be communication between lower-level clients and higher-level clients, to complete follow-on tasks. Hence, guaranteeing a smooth learning process depends heavily on much internal scheduling and task allocation, all of which requires communication. Because of this, the overall risk of information leaks is much higher with DML preventing information leaks over both the training data and the model parameters usually mean an expensive encryption mechanism need to be integrated into the system. By contrast, communication in federated learning is typically very low and highly efficient. The only communication allowed between the server and clients is for the global model transfers and learning updates aggregation, which does not involve any clients' sensitive data. Therefore, only local data privacy preserving mechanisms are necessary to protect training data as opposed to multiple mechanisms to cover the device and the communications. Further, it is more difficult for and adversary to perform an inference attack on a client's device than when data is in transit.
    
    \item Last but not least, federated learning requires dramatically less storage space than DML. DML frameworks involve a data manager that stores the entire training set, ready to be partitioned for parallel clients. The central server may act as this data manager, which substantially adds to the server’s storage load. Alternatively, the data could be managed by a third-party storage server. This option increases the risk of data leaks because it adds another entity into the system that could be malicious, plus there is an extra financial cost for data storage and maintenance. But, depending on the situation, relieving pressure on the central server may outweigh these downsides. Federated learning bypasses all these problems because the raw data is collected and processed by the client’s device, which reduces much of the storage load on the server. Further, the clients generate the training data from the data they collect as opposed to generating data specifically for model training. Hence, the impact of the learning process on the server is also drastically reduced.

\end{enumerate}

Overall, federated learning is considered to be an improved version of DML that provides substantially better privacy preservation and communications security. The vulnerabilities to adversaries are largely reduced to inference attacks as the learning results are sent to the aggregator for averaging or when learning updates are sent to the central server. Also, because the global model is publicly accessible to each participant in federated learning, adversaries can relatively easily reconstruct accurate model parameters from a client’s updates. Further, federated learning’s performance is strongly related to the update aggregation process, where each client’s contribution to the global model is the same. Consequently, just one malicious client can have a huge effect on the system. The lesson is that it is not possible to guarantee privacy and security with a framework alone. Additional privacy preserving and security mechanisms must be filled into the framework to guarantee these protections.

\section{Privacy Preservation and Security in Federated Learning}
As mentioned above, privacy is one of the paramount properties in the federated learning framework. As such, many different privacy preserving methods and security models are available to provide a meaningful privacy guarantee. In this section, we survey these methods and models and explore how each protects the various attack points in a federated learning system.

\subsection{Privacy in Federated Learning}
\subsubsection{Differential Privacy}

DP is a provable privacy concept conceived by Dwork et al. \cite{dwork2006calibrating}.  Its premise that the outputs of the queries on neighboring datasets should be statistically similar is one of the strongest standards for a privacy guarantee. Traditional DP is centralized. The formal definition of DP is presented as follows:

\begin{definition}[$\epsilon$-Differential Privacy]\label{Def:DP}
A randomized algorithm $\mathcal{M}$ gives $\epsilon$-differential privacy for any pair of
\emph{neighbouring datasets} $D$ and $D\prime$,
and for every set of outcomes $\Omega$,
$\mathcal{M}$ satisfies:
\begin{equation}
Pr[\mathcal{M}(D) \in \Omega] \leq \exp(\epsilon) \cdot Pr[\mathcal{M}(D\prime) \in \Omega]\ .
\end{equation}
\end{definition}

$\epsilon$ is the privacy parameter, also known as the \emph{privacy budget} \cite{dwork2011firm}.
It controls the level of privacy preservation. A smaller $\epsilon$ means greater privacy. \emph{Laplace} and \emph{Gaussian} mechanisms are widely used for numeric outputs of queries. The methods of DP involve adding noise to the data to obscure certain sensitive attributes until others cannot distinguish the exact true answers of quarries.

Ostensibly, DP involves adding noise to the data to obscure certain sensitive attributes until others cannot exactly determine the true answer to any query. However, the privacy guarantee DP offers is based on the assumption of a trustworthy data curator; DP cannot protect sensitive information from malicious data collectors or curators. Hence, to address situations where an individual needs to disclose their personal information to an untrusted curator, global DP was extended into local DP. Local DP is an improved DP model with the added restriction that an adversary is unable to learn too much sensitive information of any individual data contributor in the database. In local DP, only the owner of the data can obtain the original information, which provides strong privacy protection for individuals. A formal definition of local DP follows.

\begin{definition}[\emph{$\epsilon$-Local Differential Privacy}]\label{Def-LDP}
A randomized algorithm $\mathcal{M}$ satisfies $\epsilon$-differential privacy where $\epsilon$ is a privacy parameter and $\epsilon \geq 0$, if and only if for two inputs $d$ and $d^ \prime$, and any possible output $\Omega$ of $\mathcal{M}$, we have
\begin{equation}
Pr[\mathcal{M}(x) = \Omega] \leq e^\epsilon \cdot Pr[\mathcal{M}(x\prime) = \Omega]\ 
\end{equation}\label{eq-LDP}
\end{definition}

The main difference between local and global (traditional) DP is that, in global DP, the randomized noise is from the outputs of an algorithm over all users’ data, whereas with local DP, the randomized noise is over a single user’s data. Further, the data collector can only receive perturbed data $x\prime$ not the original data $x$, and cannot distinguish the real data $x$ and $x\prime$ with much confidence, regardless of the background knowledge. Thus, the user is given a privacy guarantee without the need for a trustworthy third party. If multiple DP algorithms are applied to the same dataset, the total privacy level equals the sum of the privacy budget of each algorithm.

\subsubsection{Differential Privacy in Federated Learning}
\paragraph{Global Differential Privacy}

DP is a rigorous and easily-implemented privacy model that can guarantee privacy even if an adversary has auxiliary information \cite{zhu2020more}. DP has some properties that make it particularly useful for protecting privacy in deep learning: simplicity, composition ability, and correlated data guarantee \cite{abadi2016deep}. Deep learning is often computationally inefficient and, because adding noise does not increase computational complexity, DP is often incorporated into deep learning frameworks as a simple privacy preserving method. Deep learning networks typically have many layers, and the composition ability protects the information in each layer of the network, which ensures that the output from deep learning is private. However, the data used to train the deep learning network may be correlated, which can increase the chances of a privacy leak. Hence, some methods consider these correlations so as to provide a better privacy guarantee.

Global DP protects client privacy by adding noise during the aggregation process on the server-side. Clients' updates are uploaded and stored in the aggregator temporarily. These updates can be treated like a dataset where the aggregation result is the “query”, and every update is one record. The goal with global DP is, therefore, to hide every client update in the aggregation result. McMahan et al. \cite{mcmahan2016communication} were the first to consider protecting user data in the training set with DP in federated learning. They argued that DP could provide a rigorous worst-case privacy guarantee, even when the adversary had arbitrary side-information, by adding random noise to the model’s training process. However, that guarantee would come with a utility cost. In their later work \cite{konevcny2016federated}, they modified a federated learning system’s central algorithm to produce a differentially private model, i.e., a releasable model that protects the privacy of all individuals contributing updates to the trained global model. The one shortcoming of global DP in federated learning is that the sensitivity is hard to be set. Sensitivity has an enormous impact on both the privacy guarantee and the model’s performance. Yet setting the sensitivity during the aggregation process is challenging because aggregators should not be able to distinguish a particular client’s update, which may negatively impact the trade-off between privacy and the model’s utility.

\paragraph{Local Differential Privacy}

In a federated learning setting, local DP is a better solution for protecting privacy from the client’s perspective. Each client protects their own data and unadulterated learning results (i.e., model updates) with a specific randomized algorithm. Noisy updates are then uploaded to the aggregator. Abadi et al. \cite{abadi2016deep} of Google were the first to proposed deep learning with DP in 2016, which was later followed by local DP to protect the individual training process of each participant in Google’s federated learning framework. These developers created a differentially private stochastic gradient descent algorithm (SGD), a moments accountant and a hyper-parameter tuning process, which, at the time, were are new algorithmic techniques for learning. They also refined the analysis of privacy costs within DP frameworks. More specifically, the algorithm assumes a training model with the parameters $\theta$. The procedure is then to minimize its loss function $\mathcal{L}(\theta)$, compute the gradient $\mathbf{g}_t(x_i) \gets \nabla_\theta \mathcal{L}(\theta, x_i)$ for a random subset of examples at the each step of the SGD and then clip the $\ell_2$ norm of each gradient $\Bar{\mathbf{g}}_t(x_i) \gets \mathbf{g}_t(x_i) /$ max$(1, \frac{\left\|\mathbf{g}_t(x_i)\right\|_2 }{C})$ where $C$ is the clipping threshold. Noise is added while computing the average $\Tilde{\mathbf{g}}_t \gets \frac{1}{L}(\sum_i\Bar{\mathbf{g}}_t(x_i)+\mathcal{N}(0, \sigma^2 C^2\mathbf{I}))$ before taking a step in the opposite direction of this average noisy gradient. The approach offers protection against a strong adversary, even with full knowledge of the training mechanism and access to the model’s parameters.

Geyer et al. \cite{geyer2017differentially} subsequently proposed an algorithm for client-side DP but still using federated optimization with the aim of hiding the participation and contributions of clients during the training process. Balancing the trade-off between privacy loss and model performance occurs during decentralized training. Instead of protecting a single data point’s contribution from an individual client in the learning model, the algorithm is designed to protect a client’s entire dataset. Altering and approximating the federated averaging process is done with a randomized mechanism consisting of two steps: random sub-sampling and distorting. Random sub-sampling means to randomly sample a subset of clients from the total pool of participating clients to update their optimized training results on in the further calculation in each communication round. The difference between the optimized local model and the global model in this round is referred to as client $k$'s update $\Delta w^k = w^k - w_t$. Distorting is the step where Gaussian noise is added to each client’s update. Each client's update is scaled as $\Delta w^k /$ max$(1, \frac{\lVert \Delta w^k \rVert _2}{S})$ to ensure that the second norm is limited to $\forall k$, $\lVert \Delta w^k \rVert _2 < S$. Originally, the developers set the clipping bound sensitivity to $S =$ median$\{\Delta w^k \}_{k\in Z_t}$ without using a randomised mechanism to compute the median. This caused a privacy violation. 

\paragraph{Discussion}

The problem with local DP is that the total volume of noise added is much greater than with global DP, which can negatively impact the model’s utility and performance. A future research direction for DP in federated learning is to find a better trade-off between privacy and utility to provide a strong privacy guarantee while maintaining acceptable model performance. Further, DP can only provide privacy at the data level but, in federated learning, communication and aggregation are crucial to updating the global model. Therefore, to guarantee secure communication, security mechanisms need to be incorporated into the framework.

\subsection{Security in Federated learning}

Security mechanisms normally concern the security of data transmission with cryptographic algorithms and protocols. In federated learning, most of the communications surround model aggregation because all devices must upload their training updates to the aggregator for averaging. To prevent leaks of any individual’s training results, a specific protocol called “secure aggregation” encrypts the client updates at the device level before they are uploaded for aggregation. The protocol guarantees that all updates are aggregated in a secure way and that any other party can only access the cipher-text of a client’s updates – even the server. These protocols involve secret sharing schemes, secure multi-party computation and homomorphic encryption. 

\subsubsection{Preliminaries of Security Mechanisms}
\paragraph {Secret Sharing Schemes}
Secret sharing schemes are widely used in many cryptographic protocols. They involve a client with a secret, a set of $n$ parties, and a collection of subsets of those parties called the access structure \cite{beimel2011secret}. The secret sharing scheme for the collection distributes shares of the secret to these parties by a dealer according to two requirements: 1) any subset in the collection can reconstruct the secret from its shares of the secret, and any subset not in the collection cannot reveal any partial information about the secret, separately. Secret sharing is motivated by the problem of secure information storage. They have since been developed for numerous other applications in cryptography and distributed computing, such as secure multiparty computation \cite{ben2019completeness, chaum1988multiparty, cramer2000general} and threshold cryptography \cite{desmedt1991shared}. Secret sharing schemes are firstly proposed by Blakley \cite{blakley1979safeguarding} and Shamir \cite{shamir1979share}. They are a t-out-of-m scheme based on a threshold, where a threshold $t$ and the number of the secret shares $m$ that any $t$ shares from these $m$ shares can reconstruct the whole secret. Ito et al. \cite{ito1989secret} construct a secret sharing schemes for general access structures. However, a major problem with this approach is that the share size required to provide general access is exponential to the number of parties. That said, secret sharing schemes are a good way to protect client updates in federated learning because they can be partitioned into many shares, which helps with the costs and vulnerability associated with communication. Overall, the performance and efficiency of secret sharing schemes depend on a good optimization strategy.

\paragraph{Secure Multi-party Computation}
Secure multi-party computation is first proposed by Yao \cite{yao1982protocols} in 1982. This technique addresses the problem of having a set of parties calculate an agreed-on function over their private inputs such that all parties can reveal the intended output without obtaining other parties’ inputs \cite{hazay2010efficient}. The idea is that all parties’ private inputs are protected by an encryption scheme that guarantees the utility of the data for accurately answering a query function. In this sense, multi-party computation is more like a general notion of secure computation comprising a set of techniques as opposed to being a single method. Over the last decade, multi-party computation has seen impressive developments in lower-level primitives, such as oblivious transfer protocols and encryption schemes with homomorphic properties. \cite{kairouz2019advances, zhirov2013practical}. In federated learning, aggregators average the updates of multiple clients, which contain sensitive information, so multi-party computation schemes are a highly suitable approach to protecting both the clients’ updates and the aggregation process.

\paragraph{Homomorphic Encryption}
Homomorphic encryption is first suggested by Rivest et al. \cite{rivest1978data} in 1978. It is an encryption scheme that allows complex mathematical operations to be performed on cipher-text without changing the nature of the encryption. The two different types of homomorphic encryption are fully homomorphic encryption and partially homomorphic encryption. Fully homomorphic encryption supports both additive and multiplicative operations, while partially homomorphic encryption only supports one or the other \cite{gentry2009fully}. Fully homomorphic encryption is strongly recommended in federated learning, even though the cost of computation is much greater because the aggregation process involves both addition and multiplication. Also, because the central server should not be able to decrypt the client updates, a trusted third party must be involved to hold a key \cite{kairouz2019advances}, and the central server must be able to sum the client updates using only cipher-text. Homomorphic encryption exactly meets all these requirements.

\subsubsection{Secure Aggregation in Federated Learning}
Secure aggregation is a subclass of multi-party computation algorithms where a group of parties that do not trust each other each hold sensitive information and must collaborate to calculate an aggregated value. The aggregated value should not reveal any party’s information (except what it can learn from its own information). Like homomorphic encryption and secret sharing schemes, each client’s outputs are encrypted before they are shared, which guarantees a secure transit process.

\paragraph{Federated Secure Aggregation Protocol}
In late 2016, Bonawitz et al. \cite{bonawitz2016practical} propose the first secure aggregation suggested secure aggregation protocol for federated learning to protect the privacy of clients' model gradients and to guarantee that the server only learns the sum of the clients’ inputs while the users learn nothing. Later, in early 2017, Bonawitz et al. \cite{bonawitz2017practical} further developed a full version of the protocol for practical applications. A random number masks each client’s raw input to prevent direct disclosure to the central server, and each client generates a private-public key pair for each epoch of the aggregation process. Each client is allowed to combine its private key and every other client's public key, to generate a private shared key with a hash function. The hash function involves Pseudo Random Generator and Decisional Diffie-Hellman assumption to guarantee each pair of clients' private shared keys are additive inverse. Because the sum of a pair of private shared keys is zero, all clients’ masks are offset during the aggregation process, and the server can offset the effect of the masks to calculate an accurate aggregation result without needing to know any of the clients’ true inputs. 

The shortcoming of this method is that if any client disconnects after obtaining the mask, but before submitting the masked inputs to the server, the dropped mask cannot be offset in the server’s sum. Consequently, the protocol includes a secret sharing scheme to partition each client’s private key as a secret. The secret requires at least a minimum threshold number of clients to contribute shares to recover the secret. If no clients disconnect before the aggregation process, the scheme is not triggered but, if a client does disconnect, the server sends a request to the other clients to contribute their shares so as to recover the client’s private key. The server then computes and removes the mask using the private key coupled with the public keys from the contributors. However, this solution is not perfect and raises a new problem in that, if a dropped client reconnects and sends its inputs to the server after its private key has been recovered, the server can reveal the true inputs simply by removing the mask. To address this new problem, an additional random number for each client creates a second mask over the first. This second mask is also partitioned as a secret through the secret sharing scheme. For connected clients, the server only needs to recover and remove the sum of all the second masks, while the first mask still protects the inputs without any negative effect on the aggregation process. And, because the inputs of disconnected clients will not contribute to the aggregation, the second mask remains in place to protect the true inputs once the first mask has been recovered and removed. This protocol provides a strong and comprehensive guarantee of security over the aggregation process, but it is not particularly efficient as the key exchanges and secret sharing scheme each add significantly to the communication cost.

\paragraph{NIKE-based Secure Aggregation Protocol}
To address these two communications burdens, Mandal et al. \cite{mandal2018nike} propose the non-interactive key establishment protocol (NIKE) and a secure aggregation protocol based on NIKE. NIKE addresses the cost of key sharing. It comprises two non-colluding cryptographic secret service providers who independently calculate pairwise polynomial functions for each client. To generate a shared private key, each client generates a private polynomial function as a private key by multiplying the two polynomial functions. Further, each client has a unique order number assigned by the server, which is public information, and any client is allowed to generate a shared private key by placing the targeted client’s order number into their private polynomial function. Thus, there is no communication cost for generating a shared key, and the protocol guarantees that each pair of client calculations with its own private polynomial function will have the same results.  

The NIKE-based secure aggregation protocol reduces the communication costs associated with the secret sharing scheme. The method involves an $\ell$-regular network network where the server randomly divides every $\ell$ clients  into neighbor groups. Each client can only calculate private shared keys with their neighbors via the NIKE protocol. These keys are then summed and added as a mask over the true inputs. A 2-out-of-3 secret sharing scheme is applied such that each client’s private shared key combined with the targeted client’s order number is divided into 3 shares. One share is held by the client, and the other two are held by the targeted client and the server, respectively. If a client disconnects, the server only needs to ask its neighbors for the shares to reconstruct and offset the mask. Consequently, the communications costs for reconstructing a disconnected client’s mask is reduced to $\ell$ times the client’s private key instead of t times for every shared private key. Again, each client generates a double mask to protect its inputs for the same reasons as outlined above.

\paragraph{PrivFL}
Mandal and Gong \cite{mandal2019privfl} further the work of Mandel et al. with a protocol called PrivFL that involves linear and logistic regression models and oblivious prediction for federated learning settings. The two regression protocols make the model more robust to user disconnections. The protocol consists of multiple two-party shared local gradient computation mechanisms followed by a global gradient share-reconstruction protocol. Here, the two parties are the server, which holds the global model and the clients who hold the shares. The server and a single client first jointly run a shared local gradient computation protocol to securely compute two shares of the local gradients. The server then constructs one share of the global gradient with all alive clients via an aggregation protocol and a second share of the global gradient from its own local gradient shares. An additive homomorphic encryption scheme and a secure aggregation protocol with practical crypto-primitives imposed at the beginning of each learning epoch guarantee a safe environment for the training process client-side and the aggregation process server-side.

\paragraph{Discussion}
In general, federated learning incorporates DP to protect the training sets and model updates held by the clients, while secure aggregation protocols consisting of secure multi-party computation, secret sharing schemes and homomorphic encryption guarantee the security of the aggregation process. It is important to note, however, that these mechanisms only protect the data; they cannot assess or protect the validity of the training results. In other words, the privacy and security mechanisms currently available for federated learning only protect client updates, not a malicious client’s contribution to the global model. In the next section, we discuss the most common and effective attack models used to infiltrate federated learning systems.

\section{Attacks and Federated Learning}

One of the greatest advantages of federated learning compared to traditional distributed machine learning is the ability to prevent direct leaks of sensitive information by a malicious data server. However, federated learning is still vulnerable to some traditional attack models. On the client-side, adversaries can infer sensitive data in the training set from the training results. Server-side, malicious agents can negatively impact the performance of the global model because, in federated learning, client updates are simply averaged without monitoring the training data or the learning process. As such, an adversarial client that uploads a malicious update to the server for aggregation can have a substantial impact on the global model. In this section, we spell out the various attack methods used to compromise federated learning and the goals and capabilities of the adversaries for each attack.

\subsection{Adversarial Goals and Capabilities}

As mentioned above, the two broad types of attacks in federated learning are inference at the client level or performance degradation at the global level. Inference attacks seek sensitive information. Performance attacks, called poisoning attacks, have two levels of scope: \textit{untargeted attacks} and \textit{targeted attacks}. In an untargeted attack, the aim is to destroy the global model by reducing its accuracy \cite{biggio2012poisoning}. Targeted attacks aim to alter the model’s behavior on only one or a few specific tasks while maintaining acceptable performance on all other tasks \cite{bhagoji2018analyzing}.

\begin{table*}[ht]
    \centering
    \begin{tabularx}{\linewidth}{ll}
        \begin{minipage}{\textwidth}
             \includegraphics[width = \linewidth]{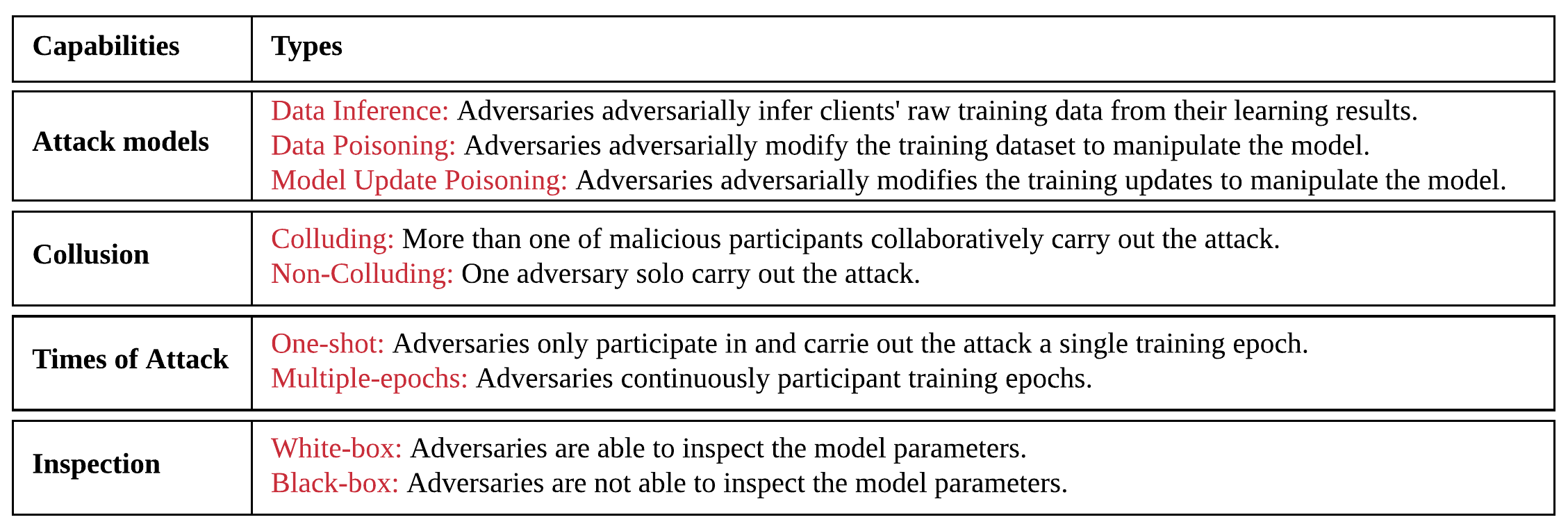}
        \end{minipage}
    \end{tabularx}
    \caption{Adversary capabilities in various attacks against federated learning}
    \label{FL-AC}
\end{table*}

\textbf{TABLE \ref{FL-AC}} summarizes the adversaries’ capabilities for each of the different types of attacks. The strategy with poisoning attacks is to act as a malicious client and upload invalid updates so as to train the model with a malicious or undesirable dataset. In most cases, poisoning attacks are executed by a solo adversary, although multiple adversaries can easily collude to attack each training epoch. Some adversaries only execute an attack once per training epoch. Further, recall that in federated learning, only a subset of all the participants is randomly chosen for each epoch, so a client may only be chosen once during the entire training process. However, when there are only a limited number of clients participating in the learning task, an adversary may be able to execute repeated attacks across multiple training epochs. The last adversarial capability is model inspections. Some models are white-boxes where the model’s parameters are ‘public’; others are black boxes where they are not. Most attacks in federated learning are white-box attacks because all clients receive the parameters of the global model.

\subsection{Inference Attack}
\subsubsection{Preliminary of Membership Inference Attack}

A membership inference attack is a common tracing attack to determine whether a specific individual is a member of a given dataset or not \cite{dwork2017exposed}. In machine learning, deep learning or federated learning, membership inference attacks aim to determine whether a given data point is present in the training data \cite{yeom2018privacy}. The accuracy of the attack corresponds to the proportion of correct membership predictions made by the adversary, while precision is calculated as the proportion of examples inferred to be members that are indeed members of the target model’s training set. These attacks take advantage of the fact that the behavior of a training model between the training set and the test set may be very different (i.e., the model may be overfit). As such, an adversary can train a machine learning model to recognize the differences in its own behavior versus the target model to determine whether or not an input record is involved in the training process \cite{yeom2018privacy, rahman2018membership, nasr2019comprehensive}.

\subsubsection{Membership Inference Attacks against Federated Learning}

Even though black-box attacks in federated learning are rare, there have been recent studies into these types of attacks in machine and deep learning scenarios. Here, the attackers can only observe the target model’s outputs as an external spectator \cite{yeom2018privacy, shokri2017membership, nasr2019comprehensive}. The findings of these studies generally show that the distribution of the training data and the generalizability of the training model are the most significant contributors to privacy leaks. Moreover, over-fit models are more susceptible to membership inference attacks. 

\paragraph{Shadow Models in Black-box Setting}

Shokri et al. \cite{shokri2017membership} proposed a membership inference attack method based on a shadow training technique. The strategy is to build many shadow models trained in a similar way to the target model with the same training algorithms (e.g. SVM, neural network) and model structure. However, while the shadow training sets have the same format as the target model, they are disjoint. Importantly, the attackers know whether or not a given record is in the training set they give to each of their shadow models. The next step is to train a neural network model using the inputs and the corresponding outputs labeled with "in" or "out" meaning in the training set or out of the training set of the shadow models. Now the attack model can distinguish between the output of the various shadow models based on memberships in the training sets. The accuracy of the attack model rises as the number of shadow models increases.

A similar idea is proposed by Rahman et al. \cite{rahman2018membership} to attack differentially private deep learning models \cite{abadi2016deep}. These researchers use two performance metrics, precision and $F_1$-score, as assessment metrics in a series of experiments designed to test the vulnerability of differentially private deep learning models. $F_1$-score is correlated to with precision and recall which is the proportion of the images belonging to the target model's training dataset that are correctly predicted to be members. The results revealed moderate vulnerability to membership inference attacks but with acceptable utility, and decreasing utility as the strength of the privacy protection grew. In other words, a model’s utility is highly correlated to the DP loss $\epsilon$.

Yeom et al. \cite{yeom2018privacy} further simplify the shadow training method by comparing the classification loss value of the target example with a preset threshold where small loss indicates membership. The idea is that this approach is equivalent to using the shadow models as a linear classifier of loss values. If the model’s confidence in an input prediction is larger than or equal to the preset threshold, it is identified as a member and a non-member otherwise. Their experiments show this strategy to be very effective with an accuracy very close to or better than the classic shadow training method. Song et al. \cite{song2019membership} follow this method but use a linear classifier for the threshold to yield a more robust deep learning model. With this approach, membership leaks are directly related to the generalization ability of the training algorithm. The more training data that is leaked, the less robust the model.

\paragraph{Deep Neural Networks in White-Box Setting}
Nasr et al. \cite{nasr2019comprehensive} recently present a comprehensive framework for analyzing data leaks with deep neural networks by executing membership inference attacks in a white-box setting. All major scenarios and adversarial capabilities in deep learning applications were considered, including model training and fine-tuning, adversaries with prior knowledge, colluding adversaries, and the vulnerabilities of SGD algorithms. A target dataset with one-hot encoding of the true labels is used to infer whether a record was included in the target model’s training set. Attackers are then able to compute the outputs of all the hidden layers, the loss, and the gradients of all layers of the target model for the given dataset. These computation results and true labels can then be used to construct the input features for an attack model consisting of convolutional neural network components and fully connected network components. Nasr and colleagues considered two roles for the attacker: first as a curious server then as a participant in a federated learning setting. A single attack model is used to process all the corresponding inputs over the observed model at once instead of running an individual independent membership inference attack on each participant’s model. Results from their experiments show that the last layer of the network leaks the most membership information. A summary of the different types of membership inference attacks in federated learning follows in \textbf{TABLE \ref{Table-Comprehensive}}.

\begin{table*}[ht]
    \centering
    \begin{tabularx}{\linewidth}{ll}
        \begin{minipage}{\linewidth}
            \includegraphics[width = \linewidth]{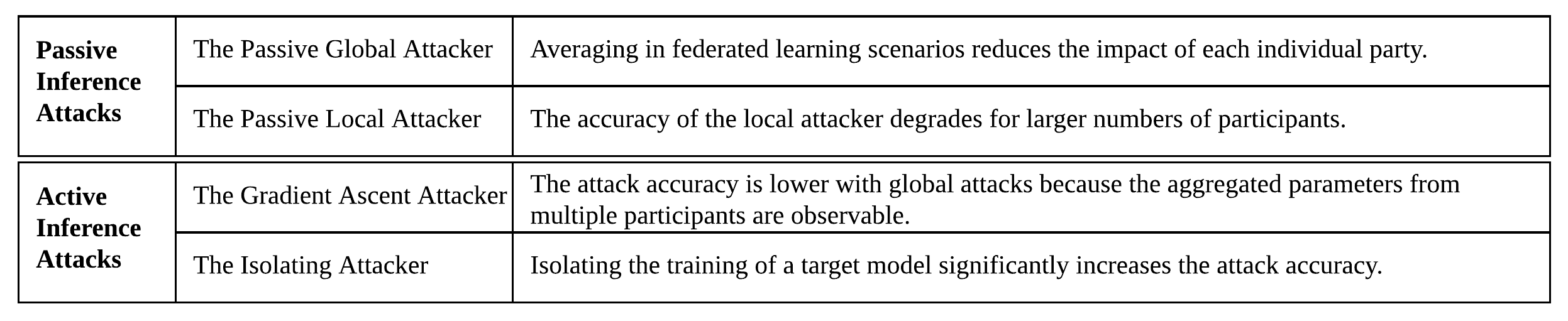}
        \end{minipage}
    \end{tabularx}
    \caption{Nasr el al. \cite{nasr2019comprehensive} Membership inference attacks in federated learning setting}
    \label{Table-Comprehensive}
\end{table*}

\subsection{Poisoning Attack}
\subsubsection{Basic Poisoning Attacks}
Poisoning attacks are a kind of causative attack \cite{biggio2012poisoning}, in which adversaries inject crafted attack points into the training data, such as manipulating a portion of the training data to adversaries' expected labels. The effect is to change the model’s parameters associated with specific learning tasks during training. Poisoned learning models subsequently misclassify those inputs at the inference stage \cite{chen2017targeted, xiao2015feature}. This attack is based on the premise that an adversary cannot directly access an existing training database, but may contribute new training data \cite{biggio2012poisoning}, which provides greater opportunities for the adversary to poison the model.

\textbf{Convergence prevention} and \textbf{backdoor attacks} are two malicious objectives of poisoning attacks \cite{baruch2019little}. The goal of a convergence prevention attack is as its name implies – to use malicious workers to ensure the model fails to converge during the training phase. Backdoor attacks are where an adversary manipulates the model during the training process so that the model produces the adversaries' expected results on an adversarial target. The target can be either a single sample or a class of samples. For instance, an adversary can make a model falsely classify either a specific person as another. The key to a good backdoor attack is to ensure the global model converges and performs well on the test set and that only the accuracy of the targeted tasks suffer.

There are two categories in data poisoning attacks: \textbf{clean-label} and \textbf{dirty-label}. In clean-label poisoning, adversaries cannot change any labels in the training data, whereas, with dirty-label poisoning, the label of a single targeted record or class can be changed. Adversaries then introduce one or more misclassified data samples into the training process. Dirty-label poisoning with deep learning models normally results in high-confidence misclassifications for the targeted record or class.

Poisoning attacks in federated learning are made possible for the following reasons \cite{bagdasaryan2018backdoor}: 
\begin{enumerate}
    \item There are usually plenty of participants in a federated learning system, which makes it easy to hide one or more malicious users in the crowd.
    \item Since all participants locally train their part of the mode, the training process is invisible to the parameter server, and the server cannot verify the authenticity of the updates \cite{hayes2018contamination}. 
    \item Local updates generated by multiple participants might be very different from each other, but the secure aggregation protocol guarantees that local updates cannot be audited by the parameter server, and the encrypted updates are simply averaged.
\end{enumerate}

\subsubsection{Poisoning Attacks Against Federated Learning}
Many recent works focus on backdoor attacks against federated learning. For example, Bagdasaryan et al. \cite{bagdasaryan2018backdoor} propose a backdoor attack with a constrain-and-scale technique where the attacker compromises one or more participants and trains a model with poisoned data. This model replace the original model as the result of federated averaging. The constrain-and-scale technique scales the model’s loss with an anomaly detection term, controlled by a hyper-parameter that the importance of evading anomaly detection. The effect is to adjust the trade-off between attack accuracy and the risk of being detected. This type of attack can be executed as either a one-shot attack or a repeated attack. In a one-shot attack setting, the accuracy of the global model on the backdoor task immediately rises to a high level in a single round when the attacker inject the backdoor updates. In a repeated attack, only a very small proportion of participants can make a better performance on the backdoor task in the target model than conventional data poisoning. 

Bhagoji et al.'s \cite{bhagoji2018analyzing} version of a poisoning attack is based on three assumptions: 1) the malicious adversary is non-colluding; 2) the data are i.i.d, which makes it easy to distinguish malicious and benign updates and harder to achieve a stealth attack; and 3) malicious adversaries have access to partial training data plus auxiliary data drawn from the same distribution as the training and test sets. The strategy is then to execute explicit boosting and alternating minimization processes. Explicit boosting overcomes the effect of scaling at the server in a gradient-based optimizer by scaling the initial updates up to $\lambda$ times, where $\lambda$ is the inverse of step rate in gradient descent. The alternating minimization mechanism boosts the part of the weight update that corresponds to the adversarial objective based on explicit boosting for malicious agents only.

Fung et al. \cite{fung2018mitigating} evaluate the vulnerability of federated learning to Sybil-based poisoning attacks without bounding the expected number of attackers and auxiliary information. The adversary performs poisoning attacks based on either the label-flipping strategy \cite{biggio2012poisoning} , in which the labels of honest training examples are flipped to an adversarial target class or the backdoor strategy \cite{bagdasaryan2018backdoor}Even with only 2 Sybils, the attack is capable of reaching a 96.2\% success rate. Fung and colleague's ultimate finding is that an adversary with enough Sybils could overpower the system regardless of the number of honest clients. Further, existing defence methods could not thwart such an attack in a federated learning setting because those methods rely on observations of the training data, and only the model parameters are observable in federated learning.

Zhang et al. \cite{zhang2019poisoning} propose a poisoning attack using generative adversarial nets (GAN) \cite{goodfellow2014generative}. Here, the adversary deploys a GAN architecture to reconstruct the private training data of other participants without invading their devices and then uses that data to train the model. First, the attacker pretends to be a benign participant to acquire training results, say under the pretext that those results are needed for a subsequent task. Simultaneously, they train a GAN to mimic prototypical samples of others. The attacker then injects an expected label into the data and generates a poisoned update. This compromises the global model's performance on the target class but not on any of the other tasks.

\paragraph{Discussion}
In general, federated learning is rather vulnerable to poisoning attacks. It is typically easy for a malicious participant to hide in the crowd of clients, and enough malicious participants can overpower the honest clients to compromise the model. Moreover, existing defense methods against such attacks are useless in federated settings because they rely on observing the training data.

Thus, there are several open questions to resolve concerning poisoning attacks. First, most existing poisoning attacks depend on repeatedly poisoning epochs rather than finding success with a one-shot approach. However, in real-world federated learning applications, very few clients are randomly selected to participate in the same training task in multiple epochs, which severely limits the chances of a successful attack with only one malicious client. Second, few poisoning attacks consider the problem of unbalanced training data. They assume that each client holds a relatively similar number of training samples, and that each client only trains one class of samples. In turn, it is assumed that both malicious clients and honest clients must contribute equally to the global model. However, it is very likely that clients will hold different numbers of samples and train a variety of classes. A challenge for adversaries is, therefore, to build an attack model capable of making a large impact on the global model with only a very small number of samples despite the large number of samples contributed by honest clients. Techniques involving data enhancement and GANs may overcome this challenge in the future.

\section{Applications Of Federated Learning}

As smartphones and Internet of Things (IoT) sensors have become ubiquitous, so too is federated learning becoming the go-to solution for scenarios involving large-scale data collection and model training, such as IoT, blockchain and recommendation systems. Compared to traditional machine learning methods, federated learning directly involves client devices in the training process as opposed to relying on a central model. As mentioned several times, offloading operations traditionally performed by the server to the end client devices gives federated learning its two key benefits: a stronger privacy guarantee and reduced communication costs.

\subsection{Federated learning in IoT Environment and Edge Computing}

\paragraph{IoT Environment and Edge Computing}
Smart homes, smart cities and smart medical systems \cite{ren2019federated, khan2019federated} are increasingly making IoT an important part of our daily life \cite{aivodji2019iotfla}. The dominant paradigm in IoT is edge computing \cite{shi2016edge}, where computation devices at the edge of the network process data downstream on behalf of cloud services and upstream on behalf of the IoT service. The result is reduced communication costs, improved data processing efficiency and better decision-making because the edge devices are both data producers and consumers. Federated learning can be thought of as an operating system for edge computing as it provides a learning protocol for coordinating the IoT devices along with all the privacy and security mechanisms and benefits outlined in Section 3 and Section 4 \cite{yang2019federated}.

\paragraph{Anomaly Detection}
Anomaly detection plays a significant role in preventing and mitigating the consequences of attacks on mobile edge networks. Many approaches to anomaly detection have been proposed for conventional machine learning and deep learning. However, the success rate of detection relies on the datasets and the sensitive information they contain \cite{lim2020federated}.Federated learning helps to address this vulnerability by storing the datasets locally. Abebe and Naveen’s \cite{abeshu2018deep} anomaly detection method for federated learning in edge networks is based on a detection model that each client helps to train using their local data. The edge nodes upload training updates to the server for aggregation, and the updated model is sent back to the device for the next training epoch. In this way, each node can improve its detection accuracy without sharing any data. A similar idea was proposed by Nguyen et al.\cite{nguyen2019diot}, where an IoT security service provider plays the role of a federated learning server, aggregating models trained by IoT gateways as clients. Both frameworks, however, assume that all edge nodes and gateways are honest and positively contribute to the training process, which means malicious participants can do significant damage.

\paragraph{Edge Caching and Computation Offloading}
Edge caching and computation offloading is another application area of federated learning, which addresses issues associated with limits to the computational power or storage capacity of the edge devices by offloading intensive computation tasks to the cloud server \cite{lim2019federated}. In these scenarios, federated learning is applied to optimize the caching and offloading among the devices. Wang et al. \cite{wang2019edge} design a near-optimal "In-Edge AI" framework using deep reinforcement learning \cite{mnih2013playing} in a federated learning setting. The method optimizes caching and offloading decisions in a mobile edge computing framework \cite{beck2014mobile} that consists of user equipment covered by base stations. Ren et al.'s \cite{ren2019federated} method is based on a similar idea of using deep reinforcement learning to optimize offloading decisions for IoT systems consisting of IoT devices and edge nodes. However, the shortcoming of both methods is that the intensity of computations in a deep reinforcement learning model’s training can cause delays in networks with a large number of heterogeneous devices. Yu et al. \cite{yu2018federated} skirt this issue with a proactive content caching scheme based on federated learning instead of deep reinforcement learning. The model optimizes the caching by making a prediction about the content’s popularity. Further, user privacy is protected by learning the parameters of the stacked auto-encoder locally without revealing the user’s personal information or content request history to the server. Similar systems of federated learning have also been applied to vehicular networks to minimize power consumption while minimizing queuing delays \cite{samarakoon2018federated} and for predicting energy demands in electric vehicle networks \cite{saputra2019energy}.

\subsection{Federated Learning in Blockchain}
\paragraph{Blockchain}

Blockchain emerged in the last decade as a way to securely transfer Bitcoin \cite{nakamoto2019bitcoin} without a central regulator. In blockchain, all user accounts and transaction information is saved in a publicly verifiable blockchain \cite{zyskind2015decentralizing}. Similar to federated learning frameworks, each client is able to access the full blockchain and locally contribute to the global blockchain by adding new blocks in chronological order \cite{swan2015blockchain}. A reward is given to a client who successfully contributes blocks to update the chain to encourage more clients to positively participate in the scheme. Client privacy is guaranteed by keeping public keys anonymous to break the flow of information back to the contributor. \cite{nakamoto2019bitcoin}. As such, all contributors are anonymous, and the public can only see that someone has added a transaction with an amount not who added the transaction.

There are several similarities between federated learning frameworks and blockchain. First, the global model (or blockchain) is accessible to every participant, and participants need to download the existing model parameters (or the chain of blocks) before participating in the update process. In federated learning, clients receive the global model from a central server, whereas, with blockchain, miners update the entire chain of blocks from broadcasts by other nodes. Second, all participants fairly contribute to the global model. Federated learning averages the client updates, while each miner in a blockchain has the same opportunity to add a new block to the chain and broadcast the update to the other miners. Third, all data processing with both systems occurs on the client device, not on a central server, and all client contributions are anonymous. In summary, these similarities lead to an appropriate combination of federated learning and blockchain for enhancing privacy and security guarantees in many existing and future applications.

\paragraph{Blockchain Empowered Federated Learning}
Lu et al. \cite{lu2019blockchain} propose a blockchain-empowered secure data sharing architecture for distributed multiple parties in an industrial IoT environment. The system comprises a permissioned blockchain module and a federated learning module. All voluntary parties who agree to share data, then upload retrieval records to the blocks of the permissioned blockchain to check if a set of queries has been proceeded. All results are learned in a federated learning setting that the multi-party data retrieval process identifies the related parties to learn the results of queries and then upload to global model instead transferring raw data directly to the data curator. The data model contains valid information towards the requests and minimized private data of participants. 

Zhao et al. \cite{zhao2019mobile} use blockchain to replace the central aggregator in a reputation-based crowdsourcing system comprising federated learning, mobile edge computing and blockchain. Clients announce their training results to a miner who checks the signature of the uploaded file. Verifiable random functions are then used to determine a subset of miners as leaders by weighting their gained rewards. These miners are preferentially responsible for averaging the client updates and uploading the global model to the blockchain once the validity of the signature has been confirmed. Only the hash of the file location is saved in the blockchain as opposed to the actual data. 

\subsection{Federated Learning in Recommendation System}

Federated learning can act as a form of privacy-preserving machine learning for recommendation systems in classical cases, such as virtual keyboard prediction \cite{hard2018federated, yang2018applied, ramaswamy2019federated} and preference recommendations \cite{hegedHus2019decentralized, ammad2019federated}. Google’s original intention with federated learning was to improve Google services on the Android system, which involve an enormous number of clients and very large-scale data. The data produced and query requests of millions of clients are simply too large to feasibly collect in a central place. As an example, an important application is Google Keyboard (Gboard), which is a virtual keyboard for mobile devices running the Android system \cite{hard2018federated}. Gboard includes typing features like next word prediction and auto-correction of spelling mistakes. It also offers expression features such as GIFs, stickers and emojis. As both a mobile application and a virtual keyboard, Gboard must guarantee its clients’ privacy because what clients type into their device can be recorded, and what is typed may be sensitive, such as passwords. Federated learning can address this problem by allowing Gboard to train a machine learning model without collecting the clients’ sensitive data \cite{yang2018applied}. Long short-term memory \cite{hochreiter1997long} is used to train a baseline model that selects and displays a query suggestion. A model is then triggered that determines if the suggestion should be shown to the client. Ramaswamy et al. \cite{ramaswamy2019federated} consider "Diversification" while focusing on emoji prediction in Gboard. A lack of diversity can lead to situations where only the most frequently-used emoji are predicted regardless of the input. To overcome this issue, Ramaswamy and colleagues scaled the probability of each emoji in keeping with an empirical probability. The proposed method is also applicable to word prediction.

Some applications keep private logs of user activities on the client’s device, such as browser histories and cookies, to help provide recommendations based on user preferences. These logs typically contain a wealth of sensitive information on a user’s interests and habits. Federated learning can protect this information from leaking given the central tenet that no data leaves the device \cite{hegedHus2019decentralized}. Ammad-ud-din et al. \cite{ammad2019federated} propose federated collaborative filtering for recommendation systems based on implicit feedback from clients. A collaborative filter trains a model to learn interactions between a client and a set of items. Then, new items that should be of interest to the client are recommended based on the learned patterns. A federated collaborative filter aggregates the gradients of each client’s filter model into the global model to make recommendations without loss of accuracy. 

\section{Conclusion}

In this paper, we surveyed federated learning in the context of data privacy preserving and security. In general, we believe that federated learning is a necessary trend in the advancement of distributed and collaborative machine learning because of its ability to offload computations from the central server. Further, federated learning accommodates the large-scale numbers of participants common to many of today’s online services in a secure way, and the privacy and security this learning framework affords is almost unparalleled. The data used to train the global never leaves the client’s device. Only the training results are uploaded to the central server as a partial model update. The received client updates are then aggregated and averaged; the global model is updated; and the server prepares the next training epoch.

The privacy of the client’s data and the model updates it transmits are protected by global DP and local DP mechanisms applied during the training process, while the global model is protected through a secure aggregation protocol consisting of secure multi-party computation protocols, secret sharing schemes and many other encryption mechanisms. However, each of these protections have specific strengths and weaknesses. For instance, global DP is vulnerable to an adversarial aggregator, whereas local DP can protect the client updates before sending them to the aggregator. However, local DP is sensitive to noise, and too much noise can impact the model’s utility. Secure aggregation protocols impose a numeric “mask” to conceal the true data used to generate the model updates while in transit that is then removed during the aggregation process to maintain accuracy. The downside here is that secure aggregation protocols are computationally expensive, and more work needs to be done to reduce their complexity.

However, federated learning is still vulnerable to data membership inference attacks and backdoor attacks. Further, adversaries are difficult to detect because federated learning usually involves a large number of participants, and each participant equally contributes to the global model. Therefore, one future direction of research is to investigate how to prevent data leaks due to inference during the training process. Another fruitful direction would be to improve the tolerance of federated learning models to anomalous updates during the aggregation process while still guaranteeing an appropriate level of utility and accuracy. The result would be more robust models.

\nocite{*}
\bibliography{wileyNJD-AMA}%

\begin{thebibliography}{10}

\bibitem{Hirt1974}
Hirt CW, Amsden AA, Cook JL. An arbitrary {L}agrangian-{E}ulerian computing
  method for all flow speeds.  {\it J {C}omput {P}hys. }1974;14(3):227--253.

\bibitem{Liska2010}
Liska R, Shashkov M, Vachal P, Wendroff B. Optimization-based synchronized
  flux-corrected conservative interpolation (remapping) of mass and momentum
  for arbitrary {L}agrangian-{E}ulerian methods.  {\it J {C}omput {P}hys.
  }2010;229(5):1467--1497.

\bibitem{Taylor1937}
Taylor GI, Green AE. Mechanism of the production of small eddies from large
  ones.  {\it P {R}oy {S}oc {L}ond {A} {M}at. }1937;158(895):499--521.
\newblock \url{https://doi.org/10.1098/rspa.1937.0036},
  \url{http://rspa.royalsocietypublishing.org/content/158/895/499}.

\bibitem{Knupp1999}
Knupp PM. Winslow smoothing on two-dimensional unstructured meshes.  {\it Eng
  {C}omput. }1999;15:263--268.

\bibitem{Kamm2000}
Kamm J. {\it Evaluation of the {S}edov-von {N}eumann-{T}aylor blast wave
  solution. } Technical {R}eport LA-UR-00-6055: Los {A}lamos {N}ational
  {L}aboratory; 2000.

\bibitem{Kucharik2003}
Kucharik M, Shashkov M, Wendroff B. An efficient linearity-and-bound-preserving
  remapping method.  {\it J {C}omput {P}hys. }2003;188(2):462--471.

\bibitem{Blanchard2015}
Blanchard G, Loubere R. {\it High-Order {C}onservative {R}emapping with a
  posteriori {MOOD} stabilization on polygonal meshes. }
  \url{https://hal.archives-ouvertes.fr/hal-01207156}, the {HAL} {O}pen
  {A}rchive, hal-01207156. Accessed January 13, 2016; 2015.

\bibitem{Burton2013}
Burton DE, Kenamond MA, Morgan NR, Carney TC, Shashkov MJ. An intersection
  based {ALE} scheme {(xALE)} for cell centered hydrodynamics {(CCH)}.  In:
  Talk at {M}ultimat 2013, {I}nternational {C}onference on {N}umerical
  {M}ethods for {M}ulti-{M}aterial {F}luid {F}lows; September 2--6, 2013; San
  {F}rancisco.
\newblock LA-UR-13-26756.2.

\bibitem{Berndt2011}
Berndt M, Breil J, Galera S, Kucharik M, Maire PH, Shashkov M. Two-step hybrid
  conservative remapping for multimaterial arbitrary {L}agrangian-{E}ulerian
  methods.  {\it J {C}omput {P}hys. }2011;230(17):6664--6687.

\bibitem{Kucharik2012}
Kucharik M, Shashkov M. One-step hybrid remapping algorithm for multi-material
  arbitrary {L}agrangian-{E}ulerian methods.  {\it J {C}omput {P}hys.
  }2012;231(7):2851--2864.

\bibitem{Breil2015}
Breil J, Alcin H, Maire PH. A swept intersection-based remapping method for
  axisymmetric {ReALE} computation.  {\it Int {J} {N}umer {M}eth {F}l.
  }2015;77(11):694--706.
\newblock Fld.3996.

\bibitem{Barth1997}
Barth TJ. Numerical methods for gasdynamic systems on unstructured meshes.  In:
   Kroner D, Rohde C, Ohlberger M, eds. {\it An {I}ntroduction to {R}ecent
  {D}evelopments in {T}heory and {N}umerics for {C}onservation {L}aws,
  {P}roceedings of the {I}nternational {S}chool on {T}heory and {N}umerics for
  {C}onservation {L}aws}, Lecture {N}otes in {C}omputational {S}cience and
  {E}ngineering. Berlin: Springer 1997.
\newblock ISBN 3-540-65081-4.

\bibitem{Lauritzen2011}
Lauritzen P, Erath C, Mittal R. On simplifying `incremental remap'-based
  transport schemes.  {\it J {C}omput {P}hys. }2011;230(22):7957--7963.

\bibitem{Klima2017}
Klima M, Kucharik M, Shashkov M. Local error analysis and comparison of the
  swept- and intersection-based remapping methods.  {\it Commun {C}omput
  {P}hys. }2017;21(2):526--558.

\bibitem{Dukowicz2000}
Dukowicz JK, Baumgardner JR. Incremental remapping as a transport/advection
  algorithm.  {\it J {C}omput {P}hys. }2000;160(1):318--335.

\bibitem{Kucharik2011}
Kucharik M, Shashkov M. Flux-based approach for conservative remap of
  multi-material quantities in {2D} arbitrary {L}agrangian-{E}ulerian
  simulations.  In:  Fo\v{r}t J, F{\"{u}}rst J, Halama J, Herbin R, Hubert F,
  eds. {\it Finite {V}olumes for {C}omplex {A}pplications {VI} {P}roblems \&
  {P}erspectives},  Springer {P}roceedings in {M}athematics, vol. 1: Springer
  2011 (pp. 623--631).

\bibitem{Kucharik2014}
Kucharik M, Shashkov M. Conservative multi-material remap for staggered
  multi-material arbitrary {L}agrangian-{E}ulerian methods.  {\it J {C}omput
  {P}hys. }2014;258:268--304.

\bibitem{Loubere2005}
Loubere R, Shashkov M. A subcell remapping method on staggered polygonal grids
  for arbitrary-{L}agrangian-{E}ulerian methods.  {\it J {C}omput {P}hys.
  }2005;209(1):105--138.

\bibitem{Caramana1998}
Caramana EJ, Shashkov MJ. Elimination of artificial grid distortion and
  hourglass-type motions by means of {L}agrangian subzonal masses and
  pressures.  {\it J {C}omput {P}hys. }1998;142(2):521--561.

\bibitem{Hoch2009}
Hoch P. {\it An arbitrary {L}agrangian-{E}ulerian strategy to solve
  compressible fluid flows. } Technical {R}eport: CEA; 2009.
\newblock HAL: hal-00366858.
  https://hal.archives-ouvertes.fr/docs/00/36/68/58/PDF/ale2d.pdf. Accessed
  January 13, 2016.

\bibitem{Shashkov1996}
Shashkov M. {\it Conservative {F}inite-{D}ifference {M}ethods on {G}eneral
  {G}rids}.
\newblock Boca Raton, Florida: CRC {P}ress; 1996.
\newblock ISBN 0-8493-7375-1.

\bibitem{Benson1992}
Benson DJ. Computational methods in {L}agrangian and {E}ulerian hydrocodes.
  {\it Comput {M}ethod {A}ppl {M}. }1992;99(2--3):235--394.

\bibitem{Margolin2003}
Margolin LG, Shashkov M. Second-order sign-preserving conservative
  interpolation (remapping) on general grids.  {\it J {C}omput {P}hys.
  }2003;184(1):266--298.

\bibitem{Kenamond2013}
Kenamond MA, Burton DE. Exact intersection remapping of multi-material
  domain-decomposed polygonal meshes.  In: Talk at {M}ultimat 2013,
  {I}nternational {C}onference on {N}umerical {M}ethods for {M}ulti-{M}aterial
  {F}luid {F}lows; September 2--6, 2013; San {F}rancisco.
\newblock LA-UR-13-26794.

\bibitem{Dukowicz1984}
Dukowicz J. Conservative rezoning (remapping) for general quadrilateral meshes.
   {\it J {C}omput {P}hys. }1984;54(3):411--424.

\bibitem{Margolin2002}
Margolin LG, Shashkov M. {\it Second-order sign-preserving remapping on general
  grids. } Technical Report LA-UR-02-525: Los {A}lamos {N}ational {L}aboratory;
  2002.

\bibitem{Mavriplis2003}
Mavriplis DJ. Revisiting the least-squares procedure for gradient
  reconstruction on unstructured meshes.  In: AIAA 2003-3986. 16th {AIAA}
  {C}omputational {F}luid {D}ynamics {C}onference; June 23--26, 2003; Orlando,
  {F}lorida.

\bibitem{Scovazzi2008}
Scovazzi G, Love E, Shashkov M. Multi-scale {L}agrangian shock hydrodynamics on
  {Q1/P0} finite elements: {T}heoretical framework and two-dimensional
  computations.  {\it Comput {M}ethod {A}ppl {M}. }2008;197(9--12):1056--1079.

\end{thebibliography}

\end{document}